\documentclass[10pt,english,journal]{IEEEtran}
\usepackage[T1]{fontenc}
\usepackage[latin9]{inputenc}
\usepackage{geometry}
\usepackage{mathptmx} 
\usepackage{amsmath}
\usepackage{xpatch}
\usepackage{soul,color}
\usepackage{amssymb}
\usepackage{esint}
\usepackage{mathtools}
\usepackage{amsthm}
\usepackage{nicefrac}
\usepackage{bigints}
\usepackage{mathtools}
\usepackage{blkarray, bigstrut}
\usepackage{physics}
\usepackage{calligra}
\usepackage{graphicx}
\usepackage{epstopdf}
\usepackage{dsfont}
\usepackage{array,ragged2e}
\usepackage[acronyms,nonumberlist,nopostdot,nomain,nogroupskip]{glossaries}
\usepackage{enumitem}
\usepackage{etoolbox}
\usepackage{babel}
\usepackage[nopar]{lipsum}
\usepackage{psfrag}
\usepackage[ruled,lined,linesnumbered]{algorithm2e}
\usepackage{algorithmic}
\usepackage{algorithm2e}
\usepackage{balance}
\usepackage{float}
\usepackage{hyperref}
\usepackage{url}
\usepackage{subcaption}

\SetKw{KwBy}{by}

\makeatletter
\newcommand{\removelatexerror}{\let\@latex@error\@gobble}
\makeatother

\graphicspath{ {Figures/} }

\xpatchcmd{\proof}{\hskip\labelsep}{\hskip5\labelsep}{}{}  
\makeatletter
\xpatchcmd{\proof}{\@addpunct{.}}{\@addpunct{:}}{}{}
\makeatother

\renewcommand\[{\begin{equation}}
\renewcommand\]{\end{equation}}
\usepackage{listings}
\usepackage{fancyvrb}
\usepackage{framed}

\usepackage{courier}
\usepackage[usenames,dvipsnames,table]{xcolor}

\definecolor{dkgreen}{rgb}{0,0.3,0}
\definecolor{gray}{rgb}{0.5,0.5,0.5}

\makeatletter
\newcommand*{\rom}[1]{\expandafter\@slowromancap\romannumeral #1@}
\makeatother

\newacronym{3gpp}{3GPP}{3rd Generation Partnership Project}
\newacronym{5g}{5G}{Fifth-Generation}
\newacronym{5gc}{5GC}{5G Core}
\newacronym{adc}{ADC}{Analog to Digital Converter}
\newacronym{afbw}{AFBW}{Average Fading Bandwidth}
\newacronym{aimd}{AIMD}{Additive Increase Multiplicative Decrease}
\newacronym{am}{AM}{Acknowledged Mode}
\newacronym{amc}{AMC}{Adaptive Modulation and Coding}
\newacronym{aoa}{AoA}{Angle of Arrival}
\newacronym{aod}{AoD}{Angle of Departure}
\newacronym{ap}{AP}{Access Point}
\newacronym{app}{APP}{Application Layer}
\newacronym{aqm}{AQM}{Active Queue Management}
\newacronym{awgn}{AGWN}{Additive White Gaussian Noise}
\newacronym{balia}{BALIA}{Balanced Link Adaptation}
\newacronym{bdp}{BDP}{Bandwidth-Delay Product}
\newacronym{ber}{BER}{Bit Error Rate}
\newacronym{bler}{BLER}{Block Error Rate}
\newacronym{bf}{BF}{Beamforming}
\newacronym{cad}{CAD}{Computer-Aided Design}
\newacronym{cbr}{CBR}{Constant Bit Rate}
\newacronym{qos}{QOS}{Quality of Service}
\newacronym{cc}{CC}{Congestion Control}
\newacronym{cdf}{$CDF$}{Cumulative Distribution Function}
\newacronym{ci}{CI}{Confidence Interval}
\newacronym{oran}{O-RAN}{Open Radio Access Network}
\newacronym{cir}{CIR}{Channel Impulse Response}
\newacronym{cn}{CN}{Core Network}
\newacronym{cp}{CP}{Control Plane}
\newacronym{cqi}{CQI}{Channel Quality Information}
\newacronym{crs}{CRS}{Cell Reference Signal}
\newacronym{csirs}{CSI-RS}{Channel State Information - Reference Signal}
\newacronym{dc}{DC}{Dual Connectivity}
\newacronym{dce}{DCE}{Direct Code Execution}
\newacronym{dci}{DCI}{Downlink Control Information}
\newacronym{llm}{LLM}{Large Language Model}
\newacronym{dl}{DL}{Downlink}
\newacronym{dmr}{DMR}{Deadline Miss Ratio}
\newacronym{dmrs}{DMRS}{DeModulation Reference Signal}
\newacronym{dray}{D-Ray}{Deterministic Ray}
\newacronym{e2e}{E2E}{End-to-End}
\newacronym{ecn}{ECN}{Explicit Congestion Notification}
\newacronym{ecdf}{ECDF}{Empirical Cumulative Distribution Function}
\newacronym{edf}{EDF}{Earliest Deadline First}
\newacronym{em}{EM}{electromagnetic}
\newacronym{enb}{eNB}{evolved Node Base}
\newacronym{endc}{EN-DC}{E-UTRAN-\gls{nr} \gls{dc}}
\newacronym{epc}{EPC}{Evolved Packet Core}
\newacronym{es}{ES}{Edge Server}
\newacronym{fdd}{FDD}{Frequency Division Duplexing}
\newacronym{fdma}{FDMA}{Frequency Division Multiple Access}
\newacronym{fray}{F-Ray}{Flashing Ray}
\newacronym{fs}{FS}{Fast Switching}
\newacronym{ftp}{FTP}{File Transfer Protocol}
\newacronym{gmm}{GMM}{Gaussian Mixture Model}
\newacronym{gnb}{gNB}{Next Generation Node Base}
\newacronym{harq}{HARQ}{Hybrid Automatic Repeat reQuest}
\newacronym{hetnet}{HetNet}{Heterogeneous Network}
\newacronym{hh}{HH}{Hard Handover}
\newacronym{hol}{HOL}{Head-of-Line}
\newacronym{hqf}{HQF}{Highest-quality-first}
\newacronym{ia}{IA}{Initial Access}
\newacronym{iab}{IAB}{Integrated Access and Backhaul}
\newacronym{ieee}{IEEE}{Institute of Electrical and Electronics Engineers}
\newacronym{imt}{IMT}{International Mobile Telecommunication}
\newacronym{inr}{INR}{Interference to Noise Ratio}
\newacronym{iot}{IoT}{Internet of Things}
\newacronym{ked}{KED}{Knife-Edge Diffraction}
\newacronym{kpi}{KPI}{Key Performance Indicator}
\newacronym{oxgpt}{GenOnet}{Generative Open xG Network Simulation}
\newacronym{ks}{KS}{Kolmogorov–Smirnov}
\newacronym{lcf}{LCF}{Level Crossing Frequency}
\newacronym{lcr}{LCR}{Level Crossing Rate}
\newacronym{los}{LoS}{Line-of-Sight}
\newacronym{lte}{LTE}{Long Term Evolution}
\newacronym{m2m}{M2M}{Machine to Machine}
\newacronym{mac}{MAC}{Medium Access Control}
\newacronym{mc}{MC}{Multi-Connectivity}
\newacronym{mcs}{MCS}{Modulation and Coding Scheme}
\newacronym{mec}{MEC}{Mobile Edge Cloud}
\newacronym{mi}{MI}{Mutual Information}
\newacronym{mib}{MIB}{Master Information Block}
\newacronym{mimo}{MIMO}{Multiple Input, Multiple Output}
\newacronym{mlr}{MLR}{Maximum-local-rate}
\newacronym[plural=\gls{mme}s,firstplural=Mobility Management Entities (MMEs)]{mme}{MME}{Mobility Management Entity}
\newacronym{mmwave}{mmWave}{millimeter wave}
\newacronym{moi}{MoI}{Method of Images}
\newacronym{mpc}{MPC}{Multi Path Component}
\newacronym{mptcp}{MPTCP}{Multipath TCP}
\newacronym{mr}{MR}{Maximum Rate}
\newacronym{mrdc}{MR-DC}{Multi \gls{rat} \gls{dc}}
\newacronym{mss}{MSS}{Maximum Segment Size}
\newacronym{mtd}{MTD}{Machine-Type Device}
\newacronym{mtu}{MTU}{Maximum Transmission Unit}
\newacronym{nfv}{NFV}{Network Function Virtualization}
\newacronym{nist}{NIST}{National Institute of Standards and Technology}
\newacronym{nlos}{NLoS}{Non-Line-of-Sight}
\newacronym{nr}{NR}{New Radio}
\newacronym{nrmse}{NRMSE}{Normalized Root Mean Square Error}
\newacronym{ns3}{ns-3}{Network Simulator 3}
\newacronym{nsa}{NSA}{Non Stand Alone}
\newacronym{o2i}{O2I}{Outdoor-to-Indoor}
\newacronym{ofdm}{OFDM}{Orthogonal Frequency Division Multiplexing}
\newacronym{pa}{PA}{Position-aware}
\newacronym{pbch}{PBCH}{Physical Broadcast Channel}
\newacronym{pdcch}{PDCCH}{Physical Downlonk Control Channel}
\newacronym{pdcp}{PDCP}{Packet Data Convergence Protocol}
\newacronym{pdsch}{PDSCH}{Physical Downlink Shared Channel}
\newacronym{pdu}{PDU}{Packet Data Unit}
\newacronym{per}{PER}{Packet Error Rate}
\newacronym{pf}{PF}{Proportional Fair}
\newacronym{pgw}{PGW}{Packet Gateway}
\newacronym{phy}{PHY}{Physical}
\newacronym{pl}{PL}{Path Loss}
\newacronym{ppp}{PPP}{Poisson Point Process}
\newacronym{prb}{PRB}{Physical Resource Block}
\newacronym{pss}{PSS}{Primary Synchronization Signal}
\newacronym{pucch}{PUCCH}{Physical Uplink Control Channel}
\newacronym{pusch}{PUSCH}{Physical Uplink Shared Channel}
\newacronym{qd}{QD}{Quasi Deterministic}
\newacronym{rach}{RACH}{Random Access Channel}
\newacronym{ran}{RAN}{Radio Access Network}
\newacronym[firstplural=Radio Access Technologies (RATs)]{rat}{RAT}{Radio Access Technology}
\newacronym{red}{RED}{Random Early Detection}
\newacronym{rf}{RF}{Radio Frequency}
\newacronym{fr}{FR}{Frequency Range}
\newacronym{rlc}{RLC}{Radio Link Control}
\newacronym{rlf}{RLF}{Radio Link Failure}
\newacronym{rr}{RR}{Round Robin}
\newacronym{rray}{R-Ray}{Random Ray}
\newacronym{rrc}{RRC}{Radio Resource Control}
\newacronym{rrm}{RRM}{Radio Resource Management}
\newacronym{rs}{RS}{Remote Server}
\newacronym{rsrp}{RSRP}{Reference Signal Received Power}
\newacronym{rsrq}{RSRQ}{Reference Signal Received Quality}
\newacronym{rss}{RSS}{Received Signal Strength}
\newacronym{rssi}{RSSI}{Received Signal Strength Indicator}
\newacronym{rt}{RT}{Ray Tracer}
\newacronym{rtt}{RTT}{Round Trip Time}
\newacronym{rw}{RW}{Receive Window}
\newacronym{rx}{RX}{Receiver}
\newacronym{sa}{SA}{standalone}
\newacronym{sack}{SACK}{Selective Acknowledgment}
\newacronym{sap}{SAP}{Service Access Point}
\newacronym{sch}{SCH}{Secondary Cell Handover}
\newacronym{scm}{SCM}{Stochastic Channel Model}
\newacronym{scoot}{SCOOT}{Split Cycle Offset Optimization Technique}
\newacronym{sdma}{SDMA}{Spatial Division Multiple Access}
\newacronym{sf}{SF}{Shadow Fading}
\newacronym{si}{SI}{Study Item}
\newacronym{sib}{SIB}{Secondary Information Block}
\newacronym{sinr}{SINR}{Signal-to-Interference-plus-Noise Ratio}
\newacronym{sir}{SIR}{Signal-to-Interference Ratio}
\newacronym{sm}{SM}{Saturation Mode}
\newacronym{snr}{SNR}{Signal-to-Noise Ratio}
\newacronym{son}{SON}{Self-Organizing Network}
\newacronym{srs}{SRS}{Sounding Reference Signal}
\newacronym{ss}{SS}{Synchronization Signal}
\newacronym{sss}{SSS}{Secondary Synchronization Signal}
\newacronym{sta}{STA}{Station}
\newacronym{svd}{SVD}{Singular Value Decomposition}
\newacronym{tb}{TB}{Transport Block}
\newacronym{tcp}{TCP}{Transmission Control Protocol}
\newacronym{udp}{UDP}{User Datagram Protocol}
\newacronym{tdd}{TDD}{Time Division Duplexing}
\newacronym{tdma}{TDMA}{Time Division Multiple Access}
\newacronym{tfl}{TfL}{Transport for London}
\newacronym{tgad}{TGad}{Task Group ad}
\newacronym{tgay}{TGay}{Task Group ay}
\newacronym{tm}{TM}{Transparent Mode}
\newacronym{trp}{TRP}{Transmitter Receiver Pair}
\newacronym{tti}{TTI}{Transmission Time Interval}
\newacronym{ttt}{TTT}{Time-to-Trigger}
\newacronym{tx}{TX}{Transmitter}
\newacronym{ue}{UEs}{User Equipments}
\newacronym{ul}{UL}{Uplink}
\newacronym{um}{UM}{Unacknowledged Mode}
\newacronym{uma}{UMa}{Urban Macro}
\newacronym{uml}{UML}{Unified Modeling Language}
\newacronym{utc}{UTC}{Urban Traffic Control}
\newacronym{vm}{VM}{Virtual Machine}
\newacronym{wbf}{WBF}{Wired Bias Function}
\newacronym{wf}{WF}{Wired-first}
\newacronym{wifi}{Wi-Fi}{Wireless Fidelity}
\newacronym{wigig}{WiGig}{Wireless Gigabit}
\newacronym{wlan}{WLAN}{Wireless Local Area Network}
\newacronym{xpr}{XPR}{Cross Polarization Ratio}
\newacronym{thz}{THz}{Terahertz}
\newacronym{ghz}{GHz}{Gigahertz}
\newacronym{ai}{AI}{Artificial Intelligence}
\newacronym{dnn}{DNN}{Deep Neural Network}
\newacronym{6g}{6G}{Sixth-Generation}
\newacronym{toa}{ToA}{Time of Arrival}
\newacronym{fsc}{FS}{Fully Stochastic}
\newacronym{hbc}{HB}{Hybrid}
\newacronym{hpbw}{HPBW}{Half Power Beamwidth}
\newacronym{tc}{TC}{Time Cluster}
\newacronym{sl}{SL}{Spatial Lobe}
\newacronym{gan}{GAN}{Generative Adversial Network}
\newacronym{cgan}{cGAN}{Conditional Generative Adversial Network}
\newacronym{relu}{$ReLU$}{Rectified Linear Unit}
\newacronym{mae}{$MAE$}{Mean Absolute Error}
\newacronym{SSCH-THz}{SSCH-THz}{Simplified Stochastic Channel Model in THz}
\newacronym{vr}{VR}{Virtual Reality}
\newacronym{edf1}{EDF}{Empirical Distribution Function}
\newacronym{sse}{SSE}{Sum Square Error}
\newacronym{soa}{SOA}{State Of Art}
\newacronym{ric}{RIC}{RAN Intelligent Controller}
\newacronym{lm}{LM}{Location Manager}
\newacronym{umi}{UMi}{urban microcells}
\newacronym{xapp}{xApp}{eXtended application}
\newacronym{ntn}{NTN}{Non-Terrestrial Network}

\usepackage{siunitx}
\usepackage{tabu}
\usepackage{booktabs}
\usepackage{multirow}
\usepackage{capt-of}
\usepackage{array}
\usepackage{arydshln}
\usepackage{cite}

\newcommand{\comment}[1]{}

\usepackage{url}
\makeatletter
\patchcmd{\@maketitle}
  {\addvspace{0.5\baselineskip}\egroup}
  {\addvspace{-1\baselineskip}\egroup}
  {}
  {}
\makeatother

\begin{document}

\title{

GenOnet: Generative Open xG Network Simulation with Multi-Agent LLM and ns-3

}

\author{\IEEEauthorblockN{
Farhad Rezazadeh\IEEEauthorrefmark{1},~Amir~Ashtari~Gargari\IEEEauthorrefmark{1},~Sandra~Lag\'en\IEEEauthorrefmark{1},~Josep~Mangues-Bafalluy\IEEEauthorrefmark{1},
Dusit~Niyato\IEEEauthorrefmark{2},
\\and Lingjia Liu\IEEEauthorrefmark{3}
}

\IEEEauthorrefmark{1}\normalsize{}Centre Tecnol\'ogic de Telecomunicacions de Catalunya (CTTC), Barcelona, Spain\\

\IEEEauthorrefmark{2}Nanyang Technological University, Singapore\\
\IEEEauthorrefmark{3}Virginia Tech, Blacksburg, USA\\

{\normalsize{}Contact Emails:  \texttt{\{name.surname\}@cttc.es},~\texttt{dniyato@ntu.edu.sg},~\texttt{ljliu@vt.edu}
}
}

\maketitle

\begin{abstract}

The move toward \gls{6g} networks relies on open interfaces and protocols for seamless interoperability across devices, vendors, and technologies. In this context, open \gls{6g} development involves multiple disciplines and requires advanced simulation approaches for testing. In this demo paper, we propose a \emph{generative simulation} approach based on a multi-agent \gls{llm} and \gls{ns3}, called \emph{\gls{oxgpt}}, to effectively generate, debug, execute, and interpret simulated Open \gls{5g} environments. The first version of \gls{oxgpt} application represents a specialized adaptation of the OpenAI GPT models. It incorporates supplementary tools, agents, 5G standards, and seamless integration with \gls{ns3} simulation capabilities, supporting both C++ variants and Python implementations. This release complies with the latest \gls{oran} and 3GPP standards.

\end{abstract}
\begin{IEEEkeywords}
Open 5G/6G, multi-agent LLM, generative simulation, ns-3
\end{IEEEkeywords}

\section{Introduction}

\IEEEPARstart{6}G focuses on implementing open interfaces and protocols to ensure smooth interoperability across various devices, vendors, and technologies~\cite{SliceOps, STEP}. In this intent, conducting a full-stack assessment of \gls{6g} cellular networks is crucial in determining the feasibility of any novel proposed approach for the next generation of wireless communication networks. \gls{6g} networks will incorporate a range of cutting-edge technologies at different levels, such as \gls{thz} communication, network management driven by \gls{ai}, Open Radio Access Network (\gls{oran}), and systems based on Non-Terrestrial Networks (NTNs). Performance bottlenecks can occur at any level of the network stack, potentially impacting the \gls{qos} for the entire system. Full-stack analysis is a method used to assess the performance and interaction of various technologies across all layers, ranging from the physical to the application layer. This analysis ensures that the technologies work together smoothly and that potential problems can be detected and resolved early. The main reason for the necessity of full-stack analysis across all layers is the unique characteristics of the underlying \gls{mmwave} and sub-\gls{thz} channels that have significant effects on the entire protocol stack~\cite{10228969}. For instance, the complexity of various essential procedures at the \gls{mac} layer, such as synchronization, control signaling, cell search, and initial access, is increased by using highly directional beams. This has an impact on both the system's robustness and delay. Another example is to validate \gls{oran} \glspl{xapp}, it is crucial to enable the analysis of \gls{oran} use cases, such as Traffic Steering (TS) for load balancing users across cells and \gls{qos} for managing bearer parameters. These scenarios should involve the utilization of interactions and patterns as multiple \gls{ue} interact with all layers of the network~\cite{Ref8ns3}.
\begin{figure}[t!]
\centering
\includegraphics[width=1\columnwidth, clip,trim={0cm 0cm 0cm 0cm}]{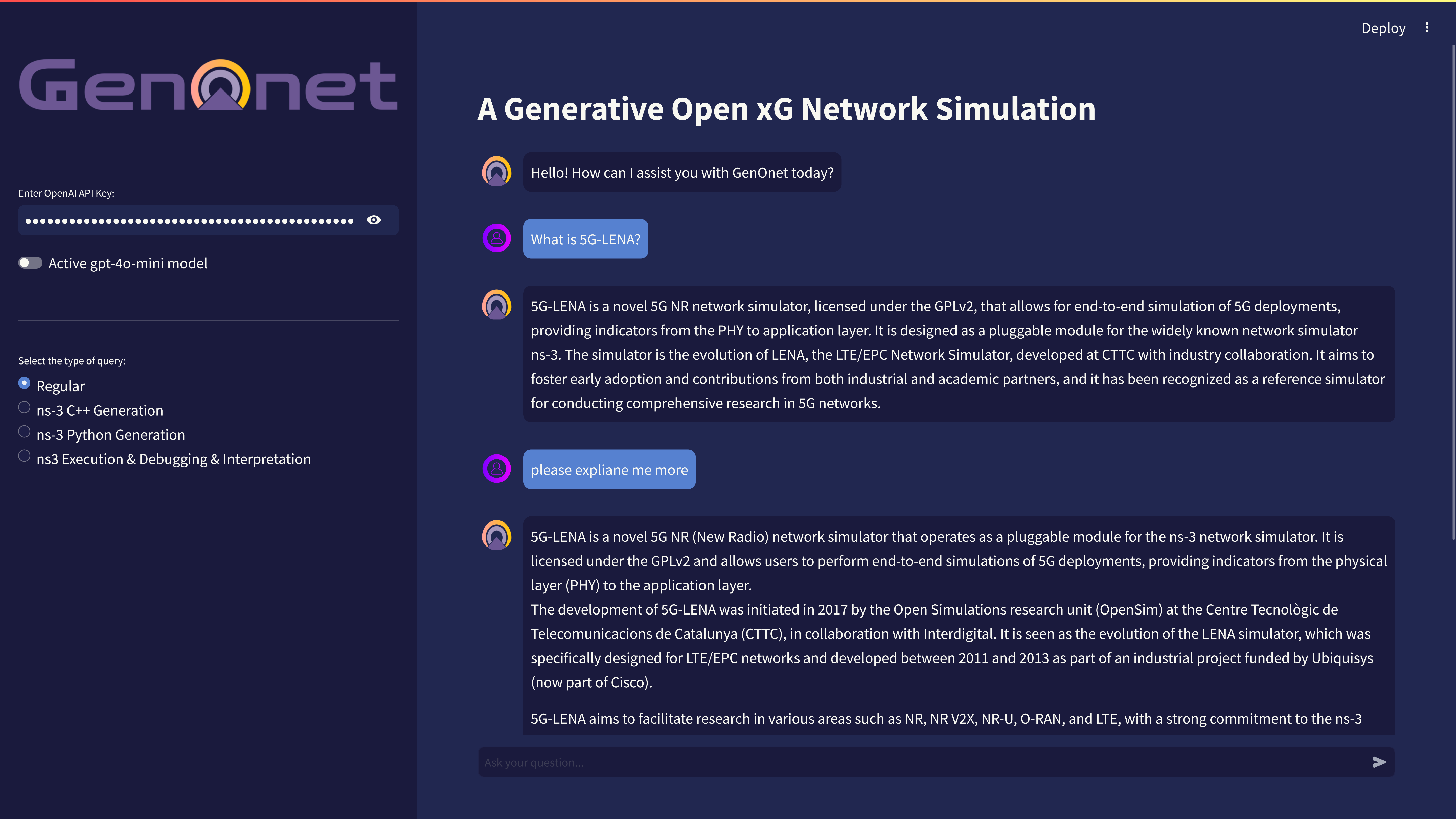}
\caption{The graphical user interface of \gls{oxgpt} application.}
\label{fig:GenOnet-GUI}
\vspace{-0.5cm}
\end{figure}

Insufficient access to testbeds for validating full-stack performance metrics can impede the confirmation of the applicability of novel proposed methods by researchers and developers for next-generation wireless communication networks. This constraint can potentially impede progress in the correct direction and mislead research and development. Discrete-event network simulators are a great alternative to evaluate performance, especially considering the limited availability of real \gls{6g} and beyond \gls{5g} (in particular \gls{fr}2 and \gls{fr}3) network deployments~\cite{fr3channel}. Discrete-event network simulators, such as \gls{ns3}\footnote{\url{https://www.nsnam.org/}}, are crucial and commonly used tools for analyzing complex networks and developing new protocols. The \gls{ns3} can accurately model several wireless and wired technologies, such as \gls{wifi} (built-in), 5G-LENA (for \gls{5g}-\gls{nr}, add-on)\footnote{\url{https://5g-lena.cttc.es/}}, and Terasim (for \gls{thz} communication, add-on)\footnote{\url{https://apps.nsnam.org/app/thz/}}, as well as the \gls{tcp}/IP protocol stack and applications. Notably, the \gls{ns3} can accurately simulate the entire network stack, encompassing all layers and applications that function within the network. This makes \gls{ns3} a great candidate for both research and industrial purposes. 

Although \gls{ns3} provides researchers and developers with features to implement and evaluate their methods comprehensively, dealing with \gls{ns3} can be challenging. The user must have extensive knowledge of all network layers, along with a proficient understanding of object-oriented programming (particularly the C++ programming language) and specific standards such as \gls{3gpp}, O-RAN, and \gls{ieee}. The combination of these skills poses many challenges to make the most of the advantages of \gls{ns3}.

Innovative approaches become essential as the importance of \glspl{llm} grows, driven by the demand for advanced agents capable of reasoning, utilizing tools, and adapting to complex real-world environments like \gls{5g}/\gls{6g} networks. We propose \gls{oxgpt} as a novel approach to address the challenges associated with the complexity of utilizing \gls{ns3} for open \gls{5g}/\gls{6g} network simulations. It leverages advanced Generative AI techniques and multi-agent \gls{llm} to automate the generation, debugging, execution, and interpretation of simulated network environments without requiring extensive programming expertise or deep knowledge of network architectures and standards. Indeed, \gls{oxgpt} effectively reduces the barriers to conducting advanced open xG network simulations. The rest of the paper is organized as follows. In Sec.~\ref{sec:system}, we describe the main features of \gls{oxgpt}. In Sec.~\ref{sec:evaluation}, we provide examples of use cases for \gls{oxgpt}, which we showcase as demonstrations. Sec.~\ref{sec:conc} concludes the work with suggestions for future research.

\section{System Overview}
\label{sec:system}
Figure~\ref{fig:GenOnet-GUI} shows the \gls{oxgpt} application's graphical user interface. This application integrates several advanced tools and models in a user-friendly application using Streamlit\footnote{\url{https://streamlit.io}}. \gls{oxgpt} emphasizes modular design using LangChain\footnote{\url{https://www.langchain.com}} and LangGraph\footnote{\url{https://www.langchain.com/langgraph}}, allowing different agents to handle specific tasks, including information retrieval based on Retrieval-Augmented Generation (RAG) technique, simulation generation, code execution, debugging, and interpretation. The questions and prompts in Figure~\ref{fig:GenOnet-GUI} illustrate that the \gls{oxgpt} framework is designed with user experience in mind, providing a smooth interface with dynamic updates and detailed feedback on the operations performed. The following is a technical analysis of how this application operates:

The \gls{oxgpt} processes queries through a chain-based sequence, where each step involves a call to an LLM, a tool, or a data preprocessing task. The technical workflow of the provided application starts with input handling, where the application receives input from the user through the chat interface. Depending on the query type (e.g., regular query, C++/Python-based ns-3 generation, or ns-3 execution), the application routes the input to the appropriate processing component. Subsequently, prompt construction occurs, where the input is used to create a detailed and context-specific prompt, leveraging templates and dynamic variables. The constructed prompt is then forwarded to the LLM during the interaction phase, where the model processes the input and generates a response. Upon receiving the response, the application performs post-processing tasks, including executing generated code, debugging, interpreting outputs, and formatting the response. Memory management is implemented through Streamlit's session state, which maintains the history of interactions, ensuring that the conversation context is preserved across multiple exchanges. Finally, in the output handling stage, the processed response is rendered in the chat interface, formatted with custom styling, and presented to the user, completing the interaction loop. This entire process is designed to operate seamlessly in real-time, providing the user with immediate feedback and dynamic updates.

\begin{figure}[t]
\centering
\includegraphics[width=1\columnwidth, clip,trim={0cm 0cm 0cm 0cm}]{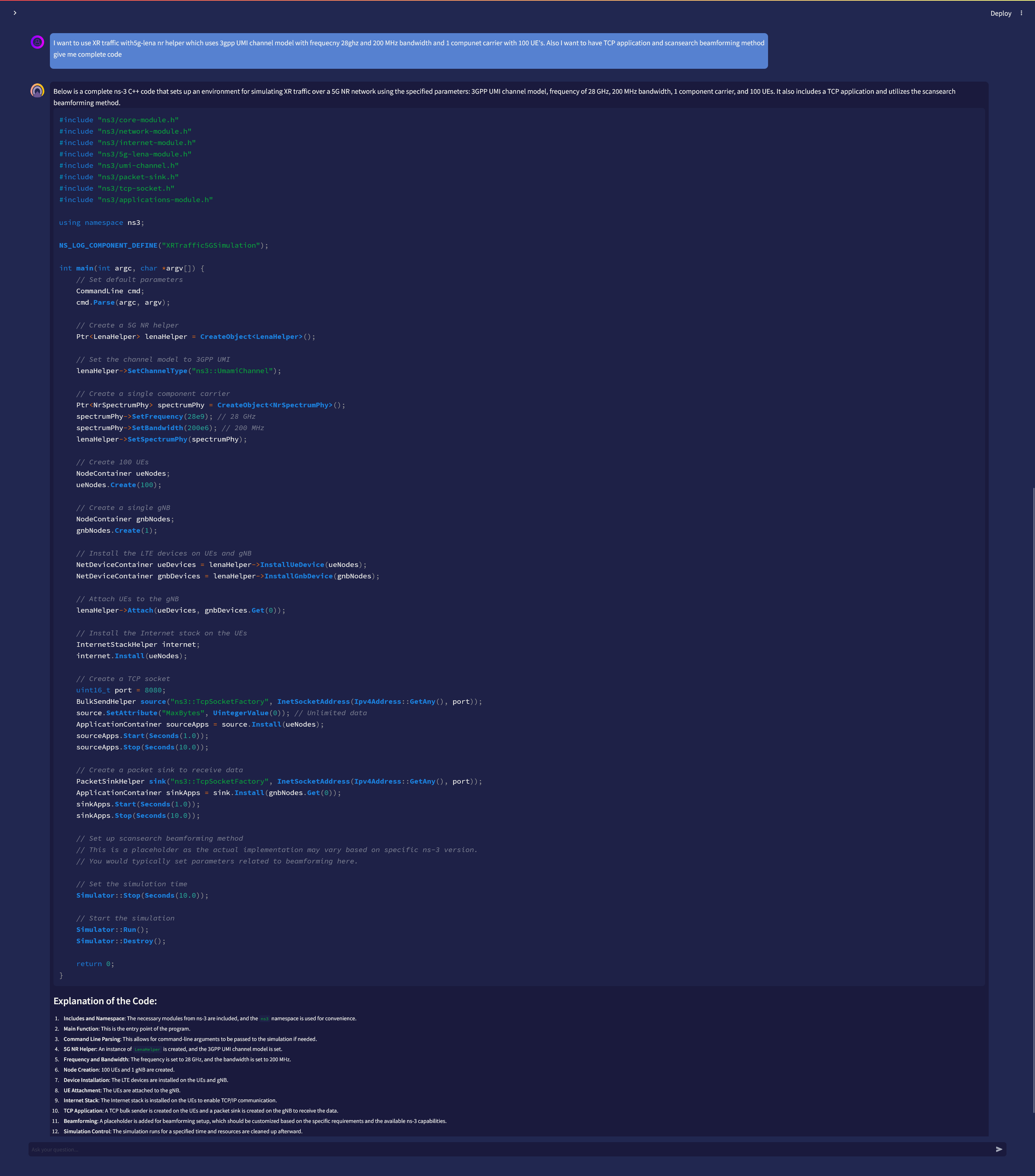}
\caption{An example of simulation generation for \gls{tcp} using the 5G-LENA \gls{nr} helper with the \gls{3gpp} standards such as the \gls{umi} channel model.}
\label{fig:Gen-5GLENA}
\end{figure}
\section{Evaluation}
\label{sec:evaluation}
Figure~\ref{fig:Gen-5GLENA} illustrates an instance of the simulation generator functionality of \gls{oxgpt}. This feature enables the app to automatically generate simulation scripts in both Python and C++ programming languages. Despite being in its initial development phase, the generator can produce ns-3 simulation scripts that offer users a thorough comprehension of the essential configuration procedures. Nevertheless, generating bug-free simulation scripts that can be compiled successfully remains challenging at this stage. Figure~\ref{fig:Gen-5GLENA} shows the result for the prompt "\textbf{\emph{I want to use XR traffic with the 5G-Lena NR helper, which uses a 3GPP UMI channel model with a frequency of 28 GHz and a 200 MHz bandwidth and 1 component carrier with 100 UE's. Also, I want to have a TCP application and a scanning beamforming method.}}" The code structure of the generator closely adheres to the ns-3 C++ code examples. It includes the required header files, the ns-3 namespace, the NS\_LOG\_COMPONENT, the use of helpers, and the simulator Run/Destroy methods. Also, the code template demonstrates how to configure channel attributes such as frequency and bandwidth based on the user's input. The quantity of gNBs and UEs, as determined by the user prompt, has been precisely configured. The code generates sample code to configure the TCP protocol using the bulkSendHelper based on the user's prompt. The code generator indicates to the users that they have to do an attachment based on \gls{5g}-Lena Helper.
\begin{figure}[t]
\centering
\includegraphics[width=1\columnwidth, clip,trim={0cm 0cm 0cm 0cm}]{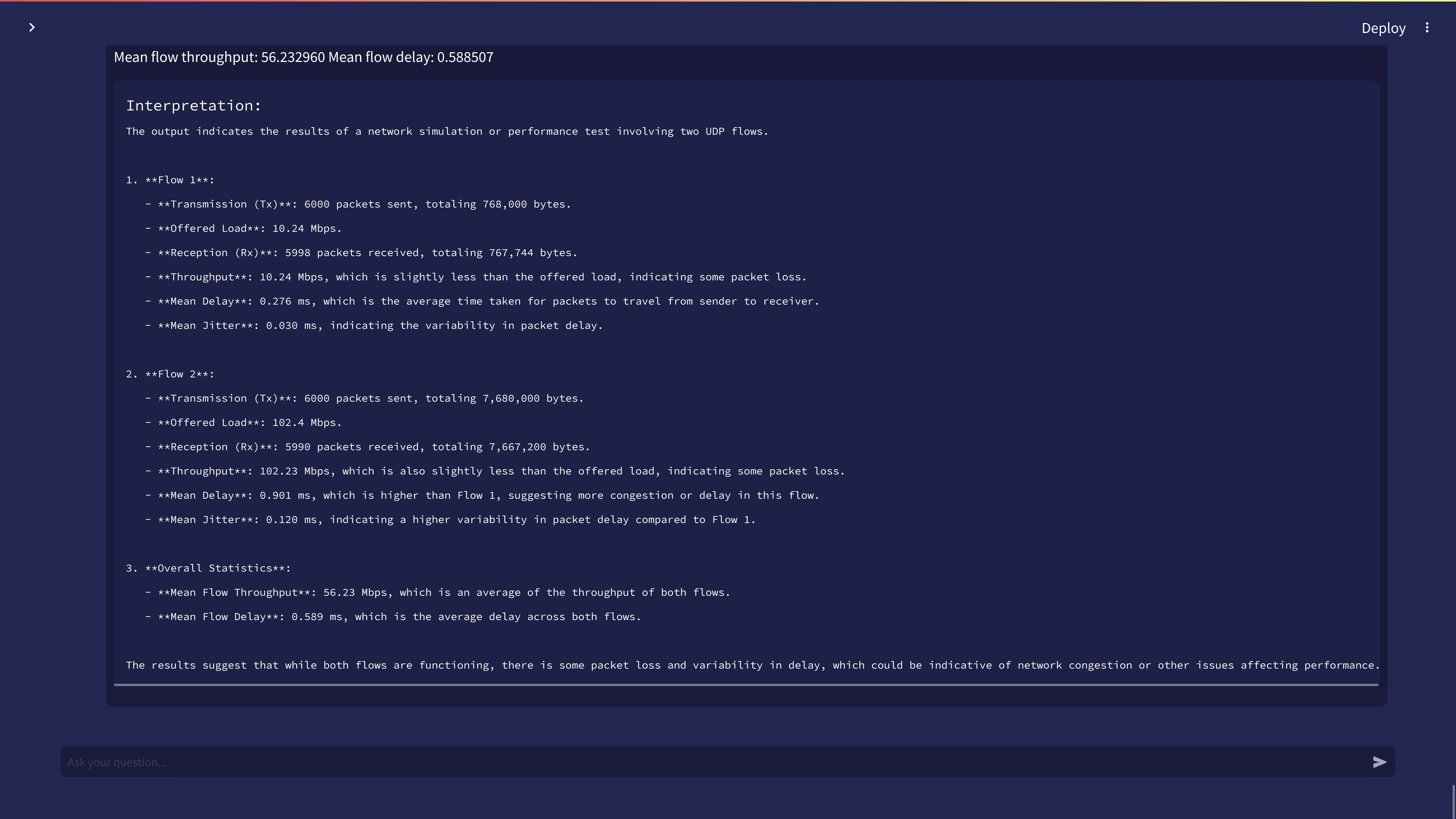}
\caption{The experimentation shows the execution and interpretation of a setup simulation using the \gls{3gpp} channel model from TR 38.901 based on the 5G-Lena \gls{nr} module.}
\label{fig:Exe-5GLENA}
\vspace{-0.5cm}
\end{figure}

Figure~\ref{fig:Exe-5GLENA}~depicts the \gls{oxgpt} app's response to running and interpreting \gls{ns3} code. This configuration utilizes the \gls{3gpp} channel model from TR 38.901\footnote{\url{https://portal.3gpp.org/desktopmodules/Specifications/SpecificationDetails.aspx?specificationId=3173}}, based on the 5G-Lena \gls{nr} module\footnote{\url{https://gitlab.com/cttc-lena/nr/-/blob/master/examples/cttc-nr-demo.cc?ref_type=heads}}. The outcomes provide comprehensive performance metrics, including throughput, delay, and jitter for two \gls{udp} flows. They showcase the effectiveness of the \gls{oxgpt} in assessing network performance across different scenarios.

\begin{figure}[t]
\centering
\includegraphics[width=1\columnwidth, clip,trim={0cm 0cm 0cm 0cm}]{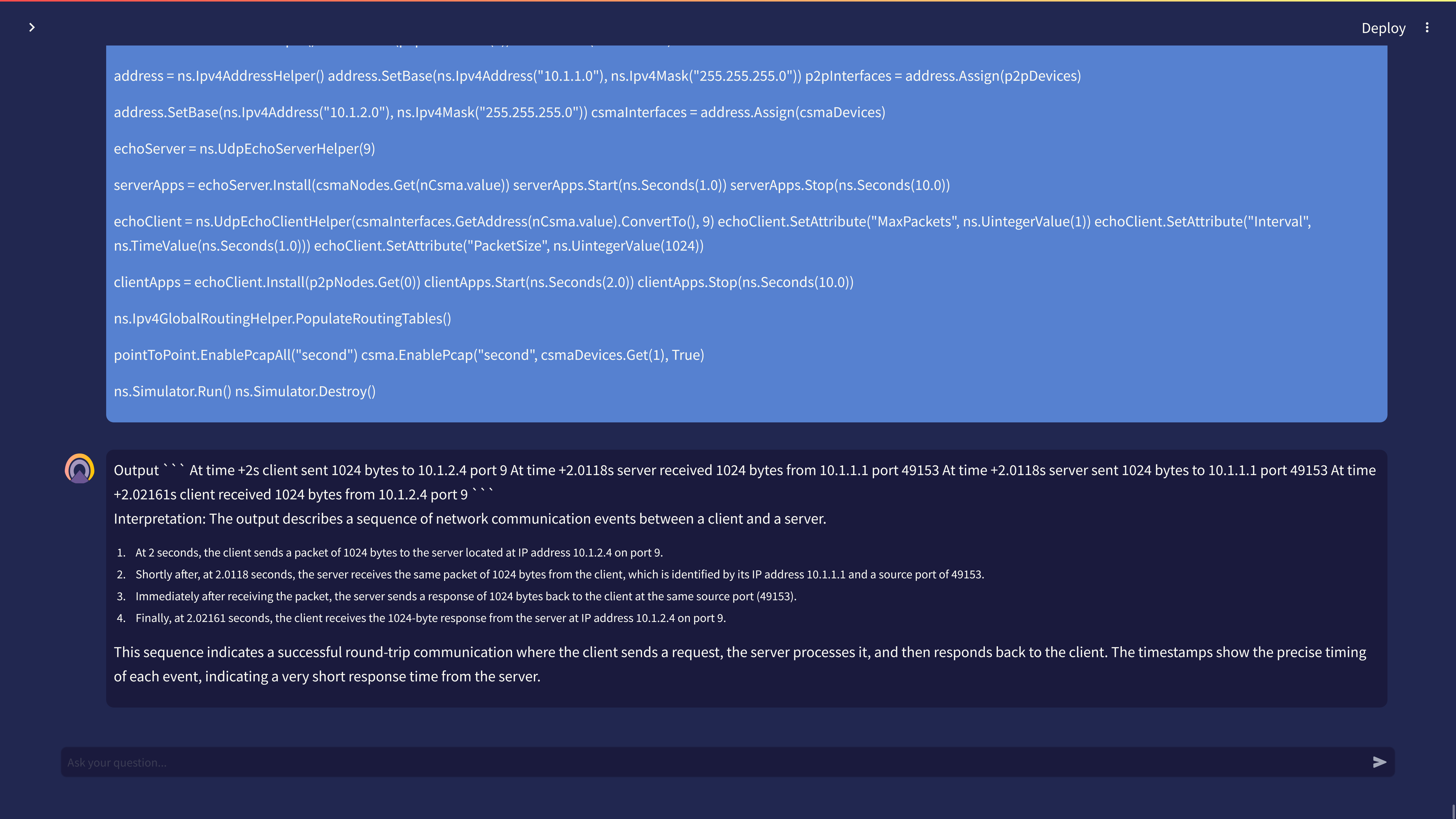}
\caption{Execution and interpretation of a Python-based \gls{ns3} example.}
\label{fig:Exe-python}
\vspace{-0.4cm}
\end{figure}

In Figure~\ref{fig:Exe-python}, we can observe a client-server communication scenario\footnote{\url{https://gitlab.com/nsnam/ns-3-dev/-/blob/master/examples/tutorial/second.py?ref_type=heads}} wherein the client transmits a 1024-byte packet to the server at time t=2 seconds. The server, located at IP address 10.1.2.4 on port 9, promptly receives the packet and sends back a response of equivalent size to the client at time t=2.0118 seconds. As explained in the interpretation, the client successfully receives the server's response at time t=2.02161 seconds, demonstrating efficient round-trip communication with precise timestamps.




\section{Conclusion and Future Work}
\label{sec:conc}
In this demo, we have presented the \gls{oxgpt} framework, a novel and innovative approach to simulating open 5G/6G network environments by leveraging multi-agent LLMs and the ns-3. It provides a flexible platform for generating, debugging, executing, and interpreting network scenarios to advance next-generation network technologies. \gls{oxgpt} integrates 5G standards and aligns with existing simulation tools, streamlining the testing and validation of open network architectures. Future developments will focus on expanding capabilities to accommodate full 5G/6G network simulations, including emerging standards and technologies. This will involve enhancements to the multi-agent LLM framework and integration of real-time data analytics and machine learning algorithms for adaptive and predictive network behaviors within the simulation.

\section*{Acknowledgment}
This work was partially funded by MCIN/AEI/ 10.13039/501100011033 grant PID2021-126431OB-I00 (ANEMONE), Spanish MINECO grant TSI-063000-2021-54 (6G-DAWN ELASTIC) and grant TSI-063000-2021-56 (6G-BLUR  SMART), and Generalitat de Catalunya grant 2021 SGR 00770.

\end{document}